\documentclass[aps,prb,twocolumn,superscriptaddress,showpacs]{revtex4}
\usepackage{graphicx}
\usepackage{bm}
\usepackage{epsfig}
\usepackage{ulem}
\usepackage{color}
\usepackage{mathrsfs}
\usepackage{dcolumn}
\usepackage{setspace}
\usepackage{array}
\usepackage{amsmath}
\usepackage{amssymb}
\usepackage{gensymb}
\usepackage{bibentry,natbib}
\usepackage{booktabs}
\usepackage{url}
\usepackage{hyperref}
\begin{document}
\bibliographystyle{unsrt}

\title{Noise signatures for determining chiral Majorana fermion modes}

\author{Yu-Hang Li}
\affiliation{International Center for Quantum Materials, School of
Physics, Peking University, Beijing 100871, China}
\author{Jie Liu}
\affiliation{Department of Applied Physics, School of Science, Xian Jiaotong University, Xian 710049, China}
\author{Haiwen Liu}
\affiliation{Center for Advanced Quantum Studies, Department of Physics, Beijing Normal University, Beijing 100875, China}
\author{Hua Jiang}
\email[]{jianghuaphy@suda.edu.cn}
\affiliation{College of Physics, Optoelectronics and Energy, Soochow
University, Suzhou 215006, China}
\author{Qing-Feng Sun}
\affiliation{International Center for Quantum Materials, School of Physics, Peking University, Beijing 100871, China}
\affiliation{Collaborative Innovation Center of Quantum Matter, Beijing 100871, China}
\affiliation{CAS Center for Excellence in Topological Quantum Computation, University of Chinese Academy of Sciences, Beijing 100190, China}
\author{X. C. Xie}
\email[]{xcxie@pku.edu.cn}
\affiliation{International Center for Quantum Materials, School of Physics, Peking University, Beijing 100871, China}
\affiliation{Collaborative Innovation Center of Quantum Matter, Beijing 100871, China}
\affiliation{CAS Center for Excellence in Topological Quantum Computation, University of Chinese Academy of Sciences, Beijing 100190, China}
\begin{abstract}
The conductance measurement of a half quantized plateau in a quantum anomalous Hall insulator-superconductor structure is reported by a recent experiment [Q. L. He \textit{et al.}, Science 357, 294-299 (2017)], which suggests the existence of the chiral Majorana fermion modes.
However, such a half quantized conductance plateau may also originates from a disorder-induced metallic phase.
To identify the exact mechanism,
we study the transport properties of such a system in the presence of strong disorders. Our results show that the local current density distributions of these two mechanisms are different. In particular, the current noises measurement can be used to distinguish them with the existing  experimental setup.
\end{abstract}
\pacs{}
\maketitle

\section{Introduction}
Chiral Majorana fermions, whose antiparticles are themselves {\cite{Majorana1937}}, have attracted extensive studies in recent years for their potential application in fault-tolerant quantum computation {\cite{Elliott2015,Alicea2012,Beenakker2013,Stern2010,Simon2008,KITEAV2003,Sarma2005,Freedman2005,Lian2017}}.
Viewed as the edge states of a topological superconductor (TSC), they were widely proposed to exist in the $\nu=5/2$ quantum Hall state {\cite{Moore1991,Read2000}}, the intrinsic $p$-wave superconductor ${\mathrm{Sr}}_{2}{\mathrm{RuO}}_{4}$ {\cite{Sarma2006}}, topological insulator or the strong spin orbital coupled semiconductor-superconductor heterostructures {\cite{Fu2008,Lutchyn2010}}, iron-based superconductors {\cite{Wang20151,Xu20161,Wu20161}}, and cold atom systems {\cite{Gurarie2005,Tewari2007}}.

In 2010, Qi {\textit{et al.}} proposed that a quantum anomalous Hall (QAH) insulator proximitied to an $s$-wave superconductor is also a promising candidate to realize a TSC {\cite{Qi2010}}. To detect them experimentally, a transport measurement in a sandwich structure formed by QAH insulator-TSC-QAH insulator will find half quantized conductance plateaus (HQCPs) {\cite{Chung2011,Wang2015}}. Such a strategy was examined recently in a magnetic doped topological insulator thin film with the central region coated by an $s$-wave superconductor, and indeed the HQCPs are observed {\cite{He2017}}.

Nevertheless, since the experimental samples are strongly disordered, some works commented that these HQCPs could also come from a disorder-induced metallic phase {\cite{Chen2018,Ji2018,Huang2018,Lian2018}}. Thus, more transport measurements are needed to rule out this possibility.
Soon afterwards, Beenakker suggested that, in a similar device with only one edge of the QAH insulator covered by a type-II superconductor, the two terminal conductance of a TSC should exhibit a distinctive $\mathrm{Z_2}$ interferometry {\cite{Beenakker2017,Fu200900,Akhmerov2009}}. Unfortunately, apart from the fact that the experimental setup needs to be re-fabricated, the magnetic field required to induce quantum flux into the superconductor region is also a magnitude larger than the coercive field. Therefore, it becomes highly desirable to propose a new method to distinguish these two mechanisms.

In this paper, we study the transport properties of a QAH insulator-superconductor structure, and suggest an experiment that is able to distinguish these two different mechanisms. Using the non-equilibrium Green's function method, we demonstrate that, though both TSC phase and metallic phase can give rise to HQCPs, the local current density distributions of these two phases are quite different.
Particularly, by analyzing the current-current correlations, we show that the related current noises of the TSC phase are zero while those of a metallic phase remain finite once the HQCP appears. 
In general, though the current noises of a disordered metal may be weakened by the self-averaging effect if the system size is very large, such noise signatures should increase with decreasing the system size, while the noise of a TSC is always zero.
This distinct feature combined with the conductance measurement provides an unambiguous evidence for determining chiral Majorana fermions with the existing experimental setup.
Finally, the application of this method in realistic experimental setup is also examined.

The rest of this paper is organized as follows. In Sec.~\ref{model}, we introduce our theoretical model. In Sec.~\ref{results}, we present the detailed results. Finally, a brief summary is presented in Sec.~\ref{summary}.

\section{Theoretical model}
\label{model}
The low energy effective Hamiltonian of a magnetically doped topological insulator thin film, in the $\psi_{\bf{k}}=\begin{pmatrix}c^t_{{\bf{k}}\uparrow},&c^t_{{\bf{k}}\downarrow},&c^b_{{\bf{k}}\uparrow},&c^b_{{\bf{k}}\downarrow}\end{pmatrix}^{T}$ space, is $\mathcal{H}_{0}\left({\bf{k}}\right)=v_Fk_y\tau_{z}\otimes\sigma_{x}-v_Fk_x\tau_{z}\otimes\sigma_{y}+m\left({\bf{k}}\right)\tau_x+M_z\sigma_z$ {\cite{Yu2010}}. Here, $c^{t/b}_{{\bf{k}}\sigma}$ annihilates an electron of momentum $\bf{k}$ and spin $\sigma$ in the top (bottom) layer. $v_F$ is the Fermi velocity. $\sigma_{x,y,z}$ and $\tau_{x,y,z}$ are Pauli matrices acting on spin and layer spaces. $m\left({\bf{k}}\right)=m_0-m_1k^2$ describes the coupling between the top and bottom layers with $m_0$ the hybridization gap and $m_1$ the parabolic band component. $M_z$ represents an external magnetic field along the $z$ direction, which can drive a phase transition from a trivial insulator with the Chern number $\mathcal{C}=0$ ($\left|M_z\right|<\left|m_0\right|$) to a QAH state with $\mathcal{C}=\pm1$ ($\left|M_z\right|>\left|m_0\right|$). The detailed parameters are fixed at $v_F=m_1=1$, $m_0=-0.1$ unless otherwise specified.

\begin{figure}[htbp]
  \centering
  \includegraphics[width=0.45\textwidth]{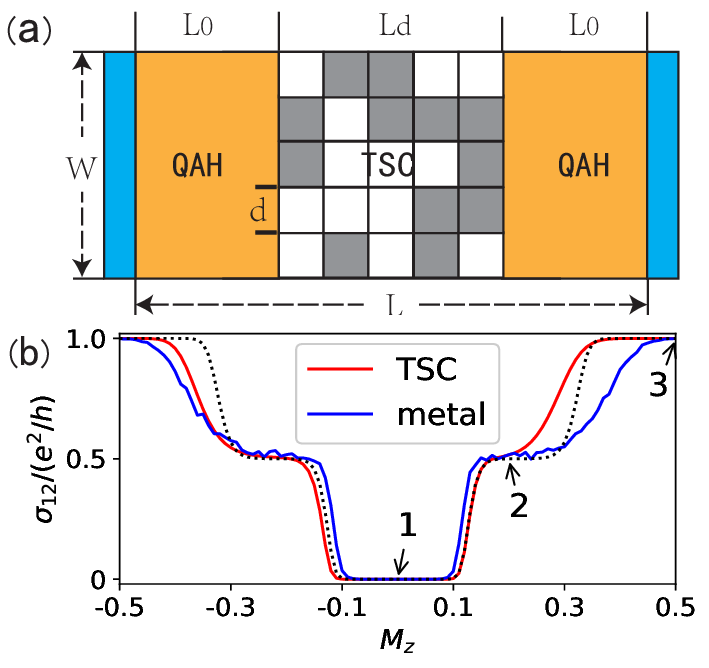}
  \caption{(color online).
  	(a) Schematic for a QAH insulator-superconductor structure with magnetic domains. A grey square denotes a random domain.
	(b) Two terminal conductance $\sigma_{12}$ as the reverse of magnetic fields. Here, the lattice unit is set as $a=1$, and the system sizes are $L_0=L_{d}=W=100$. Other parameters of the TSC are $d=20$, $w=1.0$, $\Delta_{t}=0.3$ and $\Delta_{b}=\mu=0$, while the conductance of the metal is obtained with $\Delta_{t}=\Delta_{b}=0$, $\mu=0.4$, $v=1.8$ under 1000 times average. The dash line corresponds to a clear TSC.
  	}
\label{fig1}
\end{figure}

When in proximity to an $s$-wave superconductor, the resulting Bogoliubov-de Gennes Hamiltonian in Nambu space $\Psi_{\bf{k}}=\begin{pmatrix}\psi_{\bf{k}},&\psi_{-{\bf{k}}}^{\dagger}\end{pmatrix}$ reads
\begin{equation}
\begin{split}
&\mathcal{H}_{BdG}=\left(\begin{matrix}\mathcal{H}_{0}\left({\bf{k}}\right)-\mu&{\bf{\Delta}}_{0}\\{\bf{\Delta}}_{0}^{\dagger}&-\mathcal{H}_{0}^{*}\left(-{\bf{k}}\right)+\mu\end{matrix}\right),\\
&{\bf{\Delta}}_{0}=\left(\begin{matrix}i\Delta_{t}\sigma_y&0\\0&i\Delta_{b}\sigma_y\end{matrix}\right).
\end{split}
\label{eq1}
\end{equation}
Here, $\mu$ is the chemical potential. $\Delta_{t/b}$ are the pairing potentials of the top and bottom layers, respectively.
With inequivalent $\Delta_t\ne\Delta_b$, the system could experience a series of topological phase transitions from $\mathcal{N}=0$ to $\mathcal{N}=1$ then to $\mathcal{N}=2$ with $\mathcal{N}$ being the Chern number with increasing the magnetic field strength $M_z$ {\cite{Wang2015}}. The $\mathcal{N}=1$ phase is a topological superconductor which hosts Chiral Majorana fermion modes at its edge, and the $\mathcal{N}=2$ phase is topologically equivalent to a QAH state with $\mathcal{C}=1$.

We use a percolation model to simulate the magnetic disorder as suggested in the experiment {\cite{He2017,Yasuda2017}}.
 Specifically, we divide the central region into some unit cells of size $d\times d$. Each unit cell is bound to a random domain wall ${\eta}{\,}M$ with a magnetic strength $M$ that is uniformly distributed in $\left[-w/2,w/2\right]$ and an arbitrary value ${\eta}=1$ or $0$ of equal probability as depicted in Fig.~\ref{fig1}(a).
In contrast, the metallic phase is achieved by tuning the chemical potential into the bulk band and setting the pairing potentials $\Delta_t=\Delta_b=0$ at the same time. The disorder in the metallic phase is modeled by the Anderson type random on-site potential that is uniformly distributed in $[-v/2,v/2]$.

\section{Results}
\label{results}
\subsection{Conductance}
We start by studying the half quantized two terminal conductance, which was initially believed as a fingerprint of chiral Majorana fermion modes {\cite{Chung2011,Wang2015,He2017}}. In a sandwich structure as shown in Fig.~\ref{fig1}(a),
two QAH regions experience phase transitions directly from $\mathcal{N}=-2$ to $\mathcal{N}=0$ then to $\mathcal{N}=2$ as the reversal of a magnetic field. Meanwhile, the central region experiences two additional TSC phases with $\mathcal{N}=\pm1$ near the boundaries of the phases $\mathcal{N}=\pm2$ due to the effect of superconducting proximity {\cite{Qi2010}}. In those two phases, because the normal and Andreev processes have equal probability of $1/4$, the two terminal conductance is half quantized with $\sigma_{12}=e^2/2h$ {\cite{Chung2011}}.  However, this half quantized conductance only provides a necessary condition for the  verification of chiral Majorana fermion modes. Suppose the central region is replaced with a disordered metal while the two ends remain QAH states, under a bias, an electron moving along the edge of the left QAH region is injected into the central region. After multiple scatterings by disorders inside, this electron has equal probabilities of transmission and backscattering, also resulting in a half quantized conductance {\cite{Ji2018,Chen2018,Huang2018}}.

\begin{figure*}[htbp]
  \centering
  \includegraphics[width=0.95\textwidth]{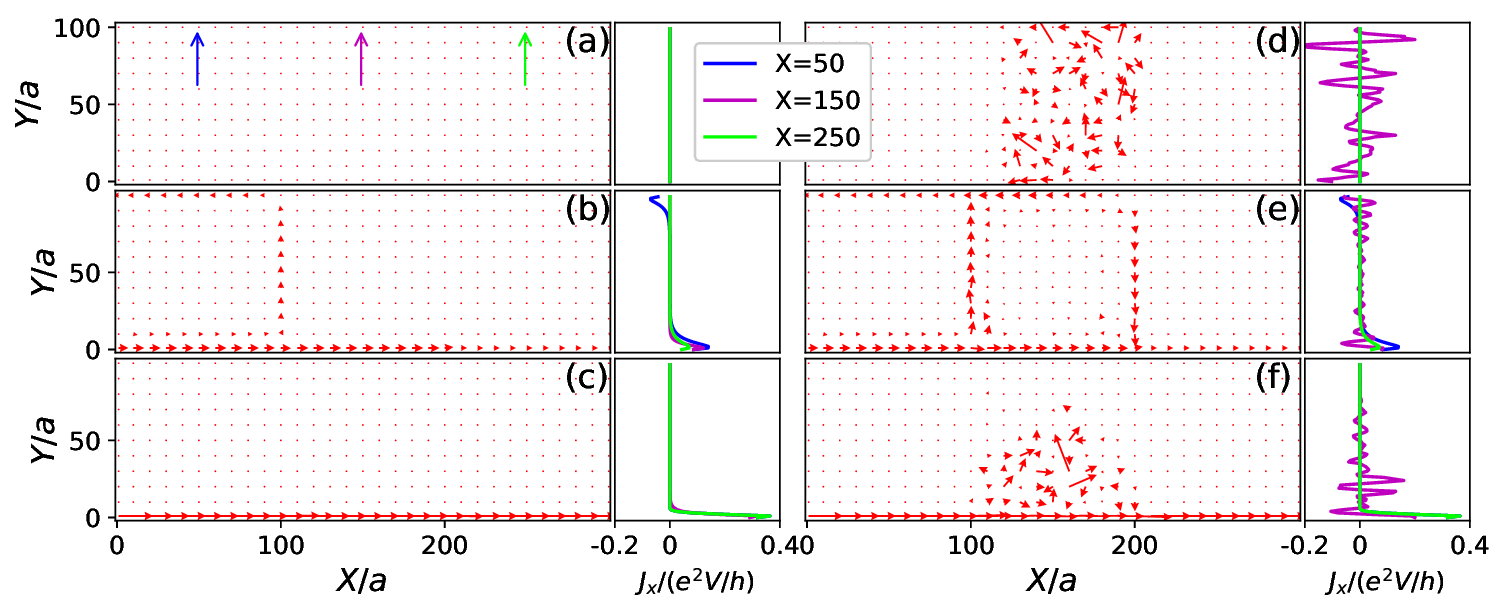}
  \caption{(color online).
  	Configurations of the local-current-flow vector for a TSC phase(left) and a metallic phase(right) at various $M_z=0$[(a),(d)], $0.2$[(b),(e)], $0.5$[(c),(f)]. The plots one-to-one correspond to three different conductance plateaus of $\sigma_{12}=0$, $0.5e^2/h$, $e^2/h$ as pointed out in Fig.~\ref{fig1}(b). The attached plots show the magnitude of the local current along the cross section at middle of three different region as labeled out by arrows of same color in (a).
  	}
\label{dis}
\end{figure*}

In Fig.~\ref{fig1}(b), the conductances $\sigma_{12}$ for both TSC phase (red line) and metallic phase (blue line) versus the external magnetic field $M_z$ are plotted. In the presence of domains, the width of the right HQCP decreases heavily while the left one remains almost unchanged. Besides, the slopes of the plateau transitions from $\mathcal{N}=\pm1$ to $\mathcal{N}=\pm2$ also decrease. Therefore, the conductance behavior is in high agreement with the experiment data reported in Ref.~[{\onlinecite{He2017}], which indicates that the percolation domains lead to an asymmetry of HQCPs.  
By contrast, the conductance of a metallic phase exhibits a similar $\sigma_{12}=e^2/2h$ plateau as shown in the same figure. When $M_{z}$ exceeds the coercive field, the left and right regions transit into QAH states with $\mathcal{N}=2$ while the central region remains a disordered metal. One can observe the same half quantized conductance due to the mechanism stated above. Enhance $M_{z}$ further, the central region also transits into a QAH state with the same Chern number, and the conductance increases to $\sigma_{12}=e^2/h$ finally. Thus, the conductance measurement is not an unambiguous evidence of chiral Majorana fermion modes.

\subsection{Local current density distributions}

Though the conductances of both TSC phase and metallic phase show the same feature, the distributions of non-equilibrium local current density from site ${\bf{i}}$ to site ${\bf{j}}$, ${\bf{J}}_{{\bf{i}}\rightarrow{\bf{j}}}$, are quite different {\cite{Jiang2009}}. In Fig.~\ref{dis}, the typical distributions of local currents for a TSC are plotted in left panels. When $M_z=0$, $\sigma_{12}=0$, all regions are trivial insulators. There is no current density in any site as shown in Fig.~\ref{dis}(a).
Tune the magnetic field strength to $M_z=0.2$, where the half quantized conductance $\sigma_{12}=e^2/2h$ appears. In this case the left and right regions are QAH insulators while the central region is generally a TSC with $\mathcal{N}=1$. The current distributions at different regions are slightly different. At left region, the currents locate on both the upper and lower edges moving on opposite directions, but at central and right regions, the currents only locate on the lower edge. However the total net currents calculated by the summation of current distributions along the $y$ direction are all $e^2/2h$, which is exactly the conductance. Enlarge the magnetic field strength further to $M_z=0.5$, all regions are QAH states with the same Chern number $\mathcal{N}=2$ and a quantized conductance $\sigma_{12}=e^2/h$ arises. 

As a comparison, the typical distributions of local currents for a disordered metal are plotted in the right panels of Fig.~\ref{dis}. 
Fig.~\ref{dis}(d) illustrates the case of $M_{z}=0$. Even though the conductance is zero, there can be some localized currents circling around disorders. If the magnetic field strength increases to $M_{z}=0.5$, the central region becomes a QAH state, but the edge currents are broaden by disorders as shown in Fig.~\ref{dis}(f).
The key difference appears at $M_z=0.2$, where the half quantized conductance arises. Different from a TSC phase, where currents only locate on the edge, the currents here spread over the entire central region as one can clearly see in Fig.~\ref{dis}(e). This indicates that the currents in a metallic phase are highly coupled.

Notably, because of the percolation domains, a TSC may also host some localized currents in the central region.
But the emergence of the HQCPs ensures that the chiral Majorana fermion modes still exist {\cite{Lian2018}}.
The direct imaging of different local currents is the most accurate way to distinguish a TSC from a metal, even though the superconductor covered in the central region makes it a challenge to image the localized currents {\cite{Lai2011,Ren2015,Nowack2013}}.
However, these different current distributions can lead to different current-current correlations, thus, give some other signatures to distinguish these two mechanisms.

\subsection{Current noises}
\begin{figure}[htbp]
  \centering
  \includegraphics[width=0.42\textwidth]{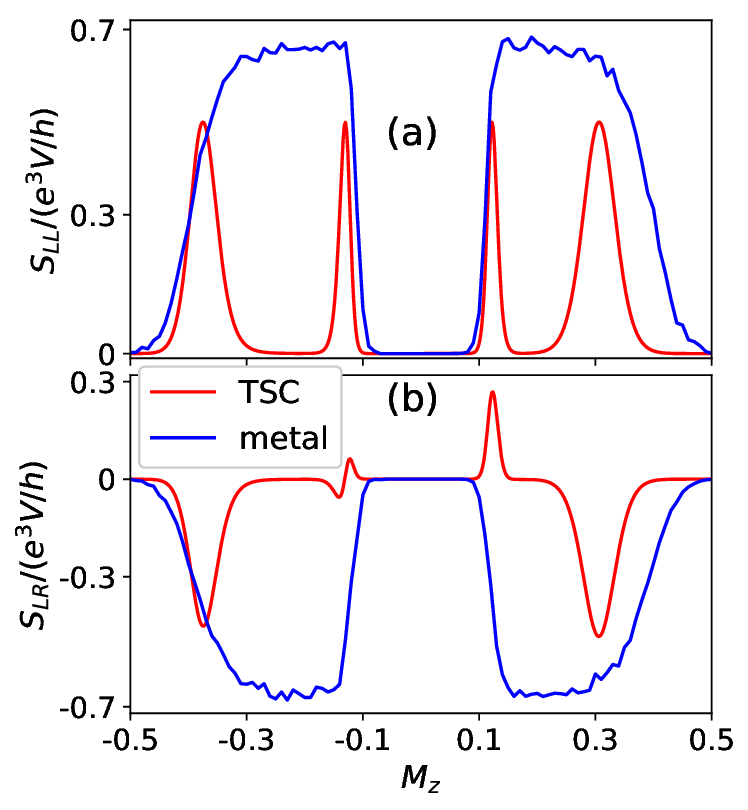}
  \caption{(color online).
  	Local (a) and non-local (b) current noises as a function of $M_z$. Here, the system parameters are taken exactly the same as Fig.~\ref{fig1}(b).
  	}
\label{noise}
\end{figure}
To analysis the current-current correlations, we study the related current noises {\cite{Blanter2000,Martin2005}}, which are defined as
\begin{equation}
S_{ab}(\omega)=\int d\tau e^{i\omega\tau}\langle\delta\hat{I}_{a}(t)\delta\hat{I}_{b}(t+\tau)+\delta\hat{I}_{b}(t+\tau)\delta\hat{I}_{a}(t)\rangle,
\label{cn}
\end{equation}
where $\delta\hat{\text{I}}\left(t\right)=\hat{\text{I}}\left(t\right)-\left<\text{I}\left(t\right)\right>$, and the subscripts $a,b$ can be $L$ or $R$ with the same (different) subscripts denoting the local (non-local) current noise. In particular, we consider the current noise at zero frequency, which represents the correlation of current fluctuations {\cite{Li2018}}.

In Fig.~\ref{noise}, both the local (a) and the non-local (b) current noises versus the magnetic field $M_z$ are plotted. The most interesting feature is that, when the conductance is half quantized, the current noises vanish if the central region is a TSC, while they remain finite otherwise. This difference can be understood in the following way.
For a metal, because electrons in the central region are randomly scattered by disorders, the current fluctuations both in the same lead and in different leads are highly coupled. Thus the noises should be visible.
 However, for a TSC since the HQCPs come from four independent Majorana fermions located at four edges of the central region, both local and non-local noises of a TSC must vanish.
Most significantly, for a TSC, as long as the HQCPs are present, these noises vanish as shown in Fig.~\ref{noise} even in the presence of strong disorders.

\subsection{Experimental setup}
Up to now, we have shown that the current noises combined with the two terminal conductances can be used as a smoking-gun evidence for the experimental confirmation of chiral Majorana fermion modes. To apply this method to a realistic experiment, the central TSC in Fig.~\ref{fig1}(a) is replaced with a magnetic doped topological insulator thin film and a coated $s$-wave superconductor {\cite{He2017}}. In the Nambu space $\Phi_{\bf{k}}=\begin{pmatrix}\phi_{\bf{k}},&\phi_{-\bf{k}}^{\dagger}\end{pmatrix}^{T}$ with $\phi_{\bf{k}}=\begin{pmatrix}c^{s}_{\bf{k}\uparrow},&c^{s}_{\bf{k}\downarrow},&c^{t}_{\bf{k}\uparrow},&c^{t}_{\bf{k}\downarrow},&c^{b}_{\bf{k}\uparrow},&c^{b}_{\bf{k}\downarrow}\end{pmatrix}^{T}$, this system can be simulated by the Hamiltonian
\begin{equation}
\mathcal{H}=\begin{pmatrix}\mathcal{H}_{c}\left(\bf{k}\right)-\mu&{\bf{\Delta}}_{c}\\ {\bf{\Delta}}_{c}^{\dagger}&-\mathcal{H}^{*}_{c}\left(-\bf{k}\right)+\mu\end{pmatrix}.
\end{equation}
Here, $\mathcal{H}_{c}\left({\bf{k}}\right)=\begin{pmatrix}\mathcal{H}_{s}\left({\bf{k}}\right)&{\bf{t}}\\{\bf{t}}^{\dagger}&\mathcal{H}_{0}\left(\bf{k}\right)\end{pmatrix}$ is the central Hamiltonian. $\mathcal{H}_{s}=\epsilon_{0}{\bf{k}}^2-\mu_{0}+\lambda\sigma_{z}$ is the Hamiltonian of a metal, where $\epsilon_{0}$ is the onsite energy, $\mu_{0}$ is the chemical potential, $\lambda$ represents the magnetic field strength.  ${\bf{\Delta}}_{c}=\begin{pmatrix}i\Delta_{s}\sigma_{y}&{\bf{0}}\\{\bf{0}}&{\bf{0}}\end{pmatrix}$ is the pairing potential, which only acts in the metal layer. The metal only couples to the top layer of the topological insulator film with the coupling matrix ${\bf{t}}=\begin{pmatrix}t_c&0&0&0\\0&t_c&0&0\end{pmatrix}$. With increasing the magnetic field strength $M_{z}$, the superconducting pairing potential satisfies the relations {\cite{Kopnin2009}}: $\lambda=0, \Delta_{s}=\Delta\left(1-M_{z}^2/M_{c}^2\right)$, when $\left|M_{z}\right|\leq M_{c}$; $\lambda=gM_{z}, \Delta_{s}=0$, otherwise. Here, $g$ is the effective land{\'e} $g$-factor, taken as $g=1$ {\cite{Zhang2015}}. $M_{c}>0$ is the critical magnetic field.

\begin{figure}[htbp]
  \centering
  \includegraphics[width=0.45\textwidth]{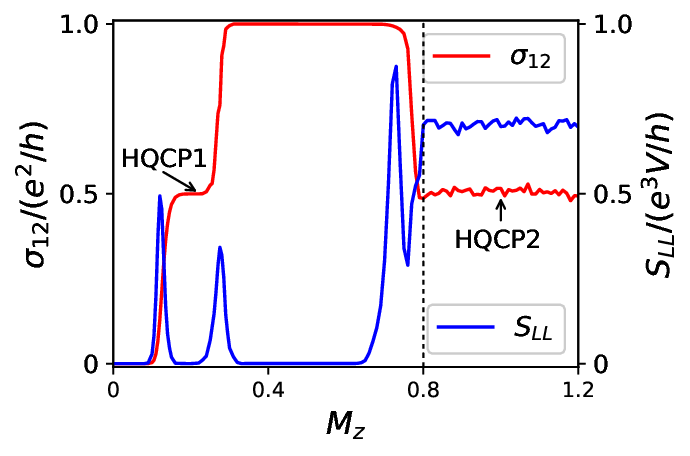}
  \caption{(color online).
  	Conductance and related current noise responses to a high magnetic field. Here, $\Delta_{s}=0.3$, $t_{c}=0.4$, $M_{c}=0.8$. The parameters in the metal layer are $\epsilon_{0}=1$, $\mu_{0}=0.5$. The disorder strength is $v=1.6$, and the disorder in the metal layer is taken as $\lambda v$ due to the Meissner effect. All datas are calculated under 1000 times average.
  	}
\label{double}
\end{figure}

Figure~\ref{double} plots the conductance and related current noise response to a high magnetic field. At low magnetic field, the conductance is zero. When the magnetic field exceeds the coercive field, the conductance is half quantized. After the magnetic field eliminates the superconducting phase, the central region becomes a disordered metal, and the conductance reaches $e^2/2h$ again. Such a conductance behavior is in perfect agreement with the experimental finding {\cite{He2017}}. Therefore, the experimental setup can be simulated by this model. To better understand the mechanisms of the two HQCPs, we perform the current noise study of this system. When HQCP1 appears, the current noise is zero. However, after the superconductor is eliminated by a high magnetic field, the existence of HQCP2 can only be attributed to a metallic phase. In this case, the current noise remains finite. This demonstrates again that the current noises can be used to determine the chiral Majorana fermion modes.

\section{Summary}
\label{summary}
In summary, we have studied the transport properties of a QAH insulator-superconductor hybrid structure. We demonstrated that both the TSC phase and the metallic phase can give arise to HQCPs but the local current distributions are different. In particular, based on a model study and a realistic simulation of the experimental situation, we show that the current noises combined with the conductance measurements can be used to distinguish these two mechanisms even though the system is strongly disordered. Our work provides an unambiguous method for determining the chiral Majorana fermion modes.

\section*{ACKNOWLEDGEMENTS} We would like to thank Xiong-Jun Liu, Qing Lin He, and especially Chui-Zhen Chen for their fruitful discussions. This work was financially supported by
National Key R and D Program of China (Grant No. 2017YFA0303301),
NBRP of China (Grant No. 2015CB921102), NSF-China (Grant Nos. 11534001 and 11574007), and
the Key Research Program of the Chinese Academy of Sciences (Grant No. XDPB08-4).
\\

\begin{appendix}
\section{Lattice models}
\label{seca}
In a square lattice $\psi_{\bf{i}}=\begin{pmatrix}c_{{\bf{i}}\uparrow}^{t},&c_{{\bf{i}}\downarrow}^{t},&c_{{\bf{i}}\uparrow}^{b},&c_{{\bf{i}}\downarrow}^{b},\end{pmatrix}^{T}$ with $c_{{\bf{i}}\sigma}^{t/b}$ annihilates an electron with spin $\sigma$ on site ${\bf{i}}$ in the top (bottom) layer, the Hamiltonian of a QAH insulator can be written as
\begin{equation}
\mathcal{H}_{0}=\sum_{\bf{i}}\left[\psi_{\bf{i}}^{\dagger}{\bf{T}}_{\bf{i}}\psi_{\bf{i}}+\left(\psi_{\bf{i}}^{\dagger}{\bf{T}}_{x}\psi_{{\bf{i}}+x}+\psi_{\bf{i}}^{\dagger}{\bf{T}}_{y}\psi_{{\bf{i}}+y}+\text{H.c.}\right)\right],
\label{QAH_lattice}
\end{equation}
with
\begin{equation*}
\begin{split}
&{\bf{T}}_{\bf{i}}=\left(m_{0}-4m_{1}\right)\tau_{x}+\left(V_{0{\bf{i}}}+M_{z}\right)\sigma_{z}, \\
&{\bf{T}}_{x}=m_{1}\tau_{x}+iv_{F}/2\tau_{z}\otimes\sigma_{y}, \\
&{\bf{T}}_{y}=m_{1}\tau_{x}-iv_{F}/2\tau_{z}\otimes\sigma_{x}.
\end{split}
\end{equation*}
Here, $V_{0{\bf{i}}}$ is an arbitary number that is uniformly distributed in $\left[-v/2, v/2\right]$, representing a magnetic disorder with strength $v$. $x$ $(y)$ is the unit vector along the $x$ $ (y)$ direction, where we have set the lattice constant $a=1$.
Other parameters have the same meaning as in the main text.
By the same token, the lattice version of the Hamiltonian of a QAH insulator coated by an ordinary metal is
\begin{equation}
\mathcal{H}_{c}=\sum_{\bf{i}}\left[\phi_{\bf{i}}^{\dagger}{\bf{D}}_{\bf{i}}\phi_{\bf{i}}+\left(\phi_{\bf{i}}^{\dagger}{\bf{D}}_{x}\phi_{{\bf{i}}+x}+\phi_{\bf{i}}^{\dagger}{\bf{D}}_{y}\phi_{{\bf{i}}+y}+\text{H.c.}\right)\right],
\end{equation}
with
\begin{equation*}
\begin{split}
&{\bf{D}}_{\bf{i}}=\begin{pmatrix}4\epsilon_{0}-\mu_{0}+\left(V_{c{\bf{i}}}+\lambda\right)\sigma_{z}&&{\bf{t}}\\{\bf{t}}^{\dagger}&&{\bf{T}}_{\bf{i}}\end{pmatrix}, \\
&{\bf{D}}_{x}=\begin{pmatrix}-\epsilon_{0}&&{\bf{0}}\\{\bf{0}}&&{\bf{T}}_{x}\end{pmatrix}, \\
&{\bf{D}}_{y}=\begin{pmatrix}-\epsilon_{0}&&{\bf{0}}\\{\bf{0}}&&{\bf{T}}_{x}\end{pmatrix}, \\
\end{split}
\end{equation*}
where $\phi_{\bf{i}}=\begin{pmatrix}c^{s}_{{\bf{i}}\uparrow},&c^{s}_{{\bf{i}}\downarrow},&c^{t}_{{\bf{i}}\uparrow},&c^{t}_{{\bf{i}}\downarrow},&c^{b}_{{\bf{i}}\uparrow},&c^{b}_{{\bf{i}}\downarrow}\end{pmatrix}^{T}$ with $c_{{\bf{i}}\sigma}^{s}$ the creating operator of an electron with spin $\sigma$ on site ${\bf{i}}$ in the metal layer, and $V_{c{\bf{i}}}=\left(\lambda/M_{z}\right)V_{0{\bf{i}}}$ represents the magnetic disorder in the metal layer.

When in proximity to an $s$-wave superconductor, the lattice version of the Bogoliubov-de Genne Hamiltonian in Nambu space is
\begin{equation}  
\mathcal{H}_{BdG}=\begin{pmatrix}H_{a}-\mu&&{\bf{\Delta}}_{a}\\{\bf{\Delta}}_{a}^{\dagger}&&-H_{a}^{*}+\mu\end{pmatrix},
\end{equation}
Here, $\mathcal{H}_{a}\left(=\mathcal{H}_{0}, \mathcal{H}_{c}\right)$ is the Hamiltonian of a QAH insulator or a QAH insulator coated by a metal. The corresponding superconducting gaps are ${\bf{\Delta}}_{0}=\left(\begin{matrix}i\Delta_{t}\sigma_y&{\bf{0}}\\{\bf{0}}&i\Delta_{b}\sigma_y\end{matrix}\right)$ and ${\bf{\Delta}}_{c}=\begin{pmatrix}i\Delta_{s}\sigma_{y}&{\bf{0}}\\{\bf{0}}&{\bf{0}}\end{pmatrix}$, respectively.

\section{Detailed derivation of the formalism}
\label{secb}
In this system, the normal tunneling $T$, normal reflection $R$, Andreev tunneling $T_{A}$, and Andreev reflection $R_{A}$ can be calculated by the scattering matrix, which has the form {\cite{Chen2017}}  
\begin{equation}
{\bf{M}}_{mn}^{\alpha\beta}=-\delta_{mn}\delta_{\alpha\beta}+i\left({\bf{\Gamma}}_{m}^{1/2}\right)^{\alpha}*\left[{\bf{G}}^{r}\right]^{\alpha\beta}*\left({\bf{\Gamma}}_{n}^{1/2}\right)^{\beta}.
\end{equation}
Here, ${\bf{G}}^{r}$ is the retarded Green's function.  ${\bf{\Gamma}}_{m}^{\alpha}=i\left[\left({\bf{\Sigma}}_{m}^{\alpha}\right)^{r}-\left({\bf{\Sigma}}_{m}^{\alpha}\right)^{a}\right]$ with $\left({\bf{\Sigma}}_{m}^{\alpha}\right)^{r}$ the retarted self energy of $\alpha\left(=e,h\right)$ particle in the $m\left(=L,R\right)$ lead.

\begin{figure}[htbp]
  \centering
  \includegraphics[width=0.45\textwidth]{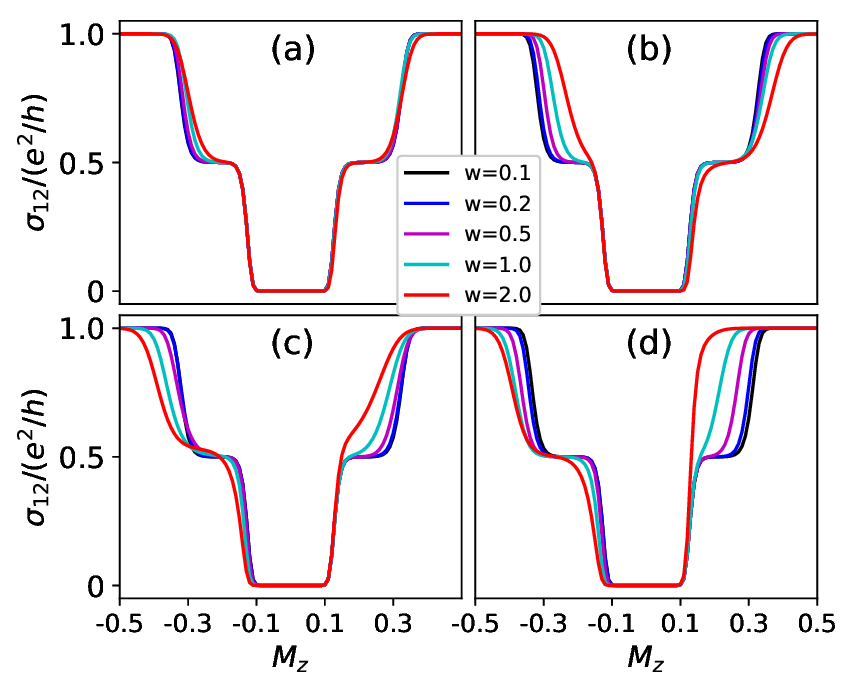}
  \caption{(color online).
             Two terminal conductance $\sigma_{12}$ versus the magnetic field as the increase of the domain strength with different domain sizes: (a) $d=5$, (b) $d=10$, (c) $d=20$, and (d) $d=25$. Other parameters are taken exactly the same as the TSC in FIG.	1 (b) in the main text.
             }
\label{a_cond}
\end{figure}

Apply a d.c. voltage $V$ on the left lead, the local current can be calculated from the evolution of the electron number operator $\hat{N}_{\bf{i}}$ in that site {\cite{Jiang2009}}
\begin{equation}
\begin{split}
{\bf{J}}_{{\bf{i}}\rightarrow{\bf{j}}}=\frac{2e}{h}&\int_{-\infty}^{0}d\epsilon\text{Im}\left\{\sigma_{z}{\bf{t}}_{\bf{ij}}\left[{\bf{G}}^{r}\left({\bf{\Gamma}}_{L}+{\bf{\Gamma}}_{R}\right){\bf{G}}^{a}\right]_{\bf{ji}}\right\}\\
&\ +\frac{2e^2}{h}\text{Im}\left[\sigma_{z}{\bf{t}}_{\bf{ij}}\left({\bf{G}}^{r}{\bf{\Gamma}}_{L}{\bf{G}}^{a}\right)_{\bf{ji}}\right]V,
\end{split}
\end{equation}
where ${\bf{t}}_{\bf{ij}}$ is the coupling Hamiltonian between site ${\bf{i}}$ and site ${\bf{j}}$. Note that we have set the right lead as the voltage ground. This current consists of two parts. The first one is the equilibrium current, while the other is the non-equilibrium current. Because the edge states of a topological nontrivial insulator is chiral, the equilibrium current moves equally along two edges with opposite directions, which does not affect the conductance.

\begin{figure}[htbp]
  \centering
  \includegraphics[width=0.45\textwidth]{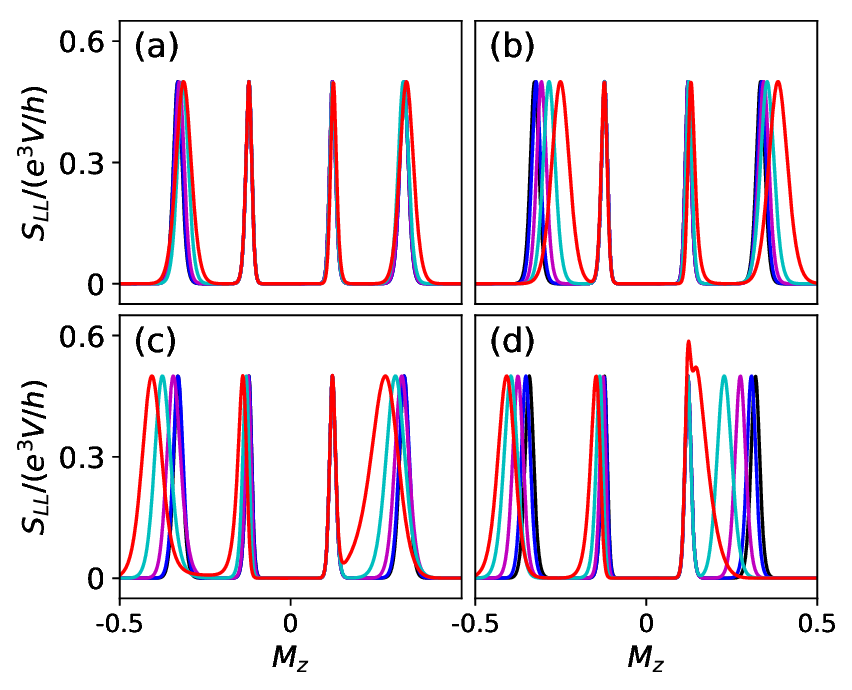}
  \caption{(color online).
             Corresponding local current noises versus the magnetic field with as the increase of the domain strength with different domain sizes: (a) $d=5$, (b) $d=10$, (c) $d=20$, and (d) $d=25$.
             }
\label{a_sll}
\end{figure}

The frequency-dependent current noises can be expressed as {\cite{Li2018}}
\begin{equation}
S_{mn}\left(\omega\right)=\frac{e^2}{\hbar}\int d\tau e^{i\omega\tau}\left[{\kappa_{mn}\left(t,t+\tau\right)+\kappa_{nm}\left(t+\tau,t\right)}\right].
\label{noise}
\end{equation}
We neglect all interactions inside the central region. With the help of the non-equilibrium Green's function method, the kernel has the form
\begin{equation*}
\begin{split}
\kappa_{mn}\left(t_{1},t_{2}\right)=
&\text{Tr}\left[\bold{t}_{Cn}\bold{G}_{nC}^{<}\left(t_{1},t_{2}\right)\bold{t}_{Cm}\bold{G}_{mC}^{>}\left(t_{2},t_{1}\right)\right.\\
&+\bold{G}_{Cm}^{<}\left(t_{1},t_{2}\right)\bold{t}_{mC}\bold{G}_{Cn}^{>}\left(t_{2},t_{1}\right)\bold{t}_{nC}\\
&-\bold{t}_{Cn}\bold{G}_{nm}^{<}\left(t_{1},t_{2}\right)\bold{t}_{mC}\bold{G}_{C}^{>}\left(t_{2},t_{1}\right)\\
&\left.-\bold{G}_{C}^{<}\left(t_{1},t_{2}\right)\bold{t}_{Cm}\bold{G}_{mn}^{>}\left(t_{2},t_{1}\right)\bold{t}_{nC}\right],
\end{split}
\label{kennel}
\end{equation*}
where $m$ and $n$ can be $L$ or $R$, representing the left or right lead. ${\bf{t}}_{C\alpha}$ and ${\bf{t}}_{\alpha C}$ $\left(\alpha=m,n\right)$ are the coupling Hamiltonian between $\alpha$ lead and the entire central region, which have the relations: ${\bf{t}}_{Cm}={\bf{t}}_{mC}^{\dagger}$. As the Green's functions depend only on the time difference, after a Fourier transformation, we obtain
\begin{equation}
S_{mn}\left(\omega\right)=\frac{e^2}{h}\int d\epsilon\left[\kappa_{mn}\left(\epsilon,\epsilon+\omega\right)+\kappa_{nm}\left(\epsilon+\omega,\epsilon\right)\right],
\end{equation}
and
\begin{equation*}
\begin{split}
\kappa_{mn}\left(\epsilon,\epsilon^{\prime}\right)=&\text{Tr}\left[\right.\bold{t}_{Cn}\bold{G}_{nC}^{<}\left(\epsilon\right)\bold{t}_{Cm}\bold{G}_{mC}^{>}\left(\epsilon^{\prime}\right)\\
&+\bold{G}_{Cm}^{<}\left(\epsilon\right)\bold{t}_{mC}\bold{G}_{Cn}^{>}\left(\epsilon^{\prime}\right)\bold{t}_{nC}\\
&-\bold{t}_{Cn}\bold{G}_{nm}^{<}\left(\epsilon\right)\bold{t}_{mC}\bold{G}_{C}^{>}\left(\epsilon^{\prime}\right)\\
&-\bold{G}_{C}^{<}\left(\epsilon\right)\bold{t}_{Cm}\bold{G}_{mn}^{>}\left(\epsilon^{\prime}\right)\bold{t}_{nC}\left.\right].
\end{split}
\end{equation*}
We couple the entire central region with the left lead first, and assume that they are equipotential. It is convenient to define ${\bf{g}}_{a}\equiv\begin{pmatrix}{\bf{g}}_{L}&&{\bf{g}}_{LC}\\{\bf{g}}_{CL}&&{\bf{g}}_{C}\end{pmatrix}$. Using the Keldysh equation ${\bf{G}}_{a}^{</>}={\bf{G}}_{a}^{r}\left({\bf{g}}_{a}^{</>}+{\bf{\Sigma}}_{a}^{</>}\right){\bf{G}}_{a}^{a}$, where the distribution self-energy is ${\bf{\Sigma}}_{a}^{</>}=\begin{pmatrix}{\bf{0}}&&{\bf{0}}\\{\bf{0}}&&{\bf{\Sigma}}_{R}^{</>}\end{pmatrix}$,  the local ($m=n$) and non-local ($m{\ne}n$) current noises can finally be written as
\begin{subequations}
\begin{equation}
\begin{split}
S_{LL}&=\frac{e^{3}V}{h}\text{Tr}\left[\right.{\bf{t}}_{CL}{\bf{G}}_{LC}^{1,<}{\bf{t}}_{CL}{\bf{G}}_{LC}^{2,>}+{\bf{G}}_{CL}^{1,<}{\bf{t}}_{LC}{\bf{G}}_{CL}^{2,>}{\bf{t}}_{LC}\\
&-{\bf{t}}_{CL}{\bf{G}}_{L}^{1,<}{\bf{t}}_{LC}{\bf{G}}_{C}^{2,>}-{\bf{G}}_{C}^{1,<}{\bf{t}}_{CL}{\bf{G}}_{L}^{2,>}{\bf{t}}_{LC}+\text{H.c.}\left.\right],
\end{split}
\end{equation}
\begin{equation}
\begin{split}
S_{LR}&=\frac{e^{3}V}{h}\text{Tr}\left[\right.{\bf{G}}_{CL}^{1,<}{\bf{t}}_{LC}\left({\bf{G}}_{C}^{r}{\bf{t}}_{CR}{\bf{g}}_{R}^{>}+{\bf{G}}_{C}^{2,>}{\bf{t}}_{CR}{\bf{g}}_{R}^{a}\right){\bf{t}}_{RC}\\
&+{\bf{t}}_{CR}{\bf{g}}_{R}^{r}{\bf{t}}_{RC}{\bf{G}}_{C}^{1,<}{\bf{t}}_{CL}{\bf{G}}_{LC}^{2,>}+\text{H.c.}\left.\right],
\end{split}
\end{equation}
\end{subequations}
where
\begin{equation*}
\begin{split}
&{\bf{G}}_{a}^{1,</>}=\begin{pmatrix}{\bf{G}}_{L}^{1,</>}&&{\bf{G}}_{LC}^{1,</>}\\{\bf{G}}_{CL}^{1,</>}&&{\bf{G}}_{C}^{1,</>}\end{pmatrix}={\bf{G}}_{a}^{r}{\bf{g}}_{a}^{</>}{\bf{G}}_{a}^{a},\\
&{\bf{G}}_{a}^{2,</>}=\begin{pmatrix}{\bf{G}}_{L}^{2,</>}&&{\bf{G}}_{LC}^{2,</>}\\{\bf{G}}_{CL}^{2,</>}&&{\bf{G}}_{C}^{2,</>}\end{pmatrix}={\bf{G}}_{a}^{r}{\bf{\Sigma}}_{a}^{</>}{\bf{G}}_{a}^{a}.
\end{split}
\end{equation*}

\begin{figure}[htbp]
  \centering
  \includegraphics[width=0.45\textwidth]{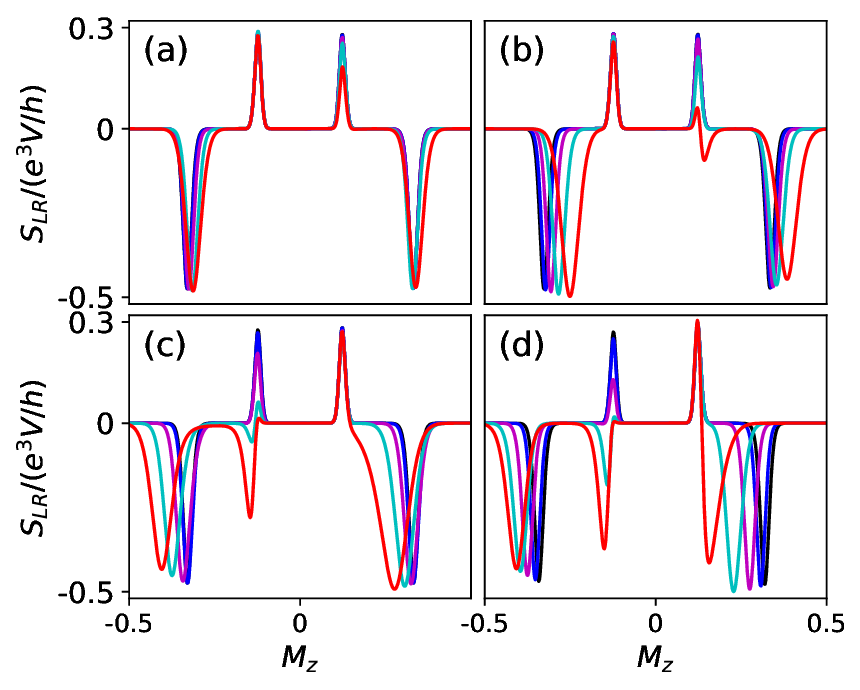}
  \caption{(color online).
             Corresponding non-local current noises versus the magnetic field with as the increase of the domain strength with different domain sizes: (a) $d=5$, (b) $d=10$, (c) $d=20$, and (d) $d=25$.
             }
\label{a_slr}
\end{figure} 

\section{Domain size}
\label{secc}

In order to better understand the effect of domains, we study the domain size dependence of the conductance. 
FIG.~\ref{a_cond} shows the two terminal conductance $\sigma_{12}$ versus the reversal of the magnetic field with different domain sizes and strengths. In the presence of domains, the edge states of the central TSC region may couple with each other through states localized at domains. Near phase boundaries, this coupling will be enhanced, which will break the quantized conductance plateaus, and meanwhile reducing the slopes of the plateau transitions. In general, this effect should manifest itself better if the domain sizes are large, which makes their coupling easily.  As it is shown in FIG.~\ref{a_cond}, the widths of HQCPs decrease heavily if we raise the domain size from (a) to (d) and disappear finally when $w=2.0$, $d=25$. Moreover, the domains also induce an asymmetry of the HQCPs, which is consistent with the experiment datas {\cite{He2017}}. This asymmetry in experimnt is widely deemed the decrease of the superconducting gap as the increase of the magnetic field. However, since the coercive field is more than one order of magnitude lower than the critical field of the superconducting metal, the decrease of the superconducting gap is negligible and could not induce such a heavy asymmetry. One possible hypothesis is that the percolation domains induce a magnetic field renormalization in the central region with $M_{z}^{\prime}=M_{z}+{\delta}M_{z}$.

If the current noises are robust to these magnetic domains? To answer this question, we present the corresponding local (FIG.~\ref{a_sll}) and non-local (FIG.~\ref{a_slr}) current noises with same parameters. The most striking feature shared by these two noises is that, the vanishing current noises appear where the conductance is quantized. Therefore,  as long as the HQCPs are not destroyed by domains, both the local and non-local current noises should disappear. Since the HQCPs have been observed in experiment, this feature indicates that, regardless of the strong disorders,  the vanishing current noises can also be observed in experiment if the HQCPs are caused by a TSC.

\end{appendix}

\hspace{3mm}


\end{document}